\documentclass[12pt]{elsart}

\usepackage{amssymb}
\usepackage[intlimits]{amsmath}
\usepackage{graphicx}
\usepackage{indentfirst}

\begin{document}

%%%%%%%%%%%%%%%%%%%%%%%%%%%%%%%%%%%%%%%%%%%%%%%%%%%%%%%%%%%%%%%%%%%%%%
%%% FRONTMATTER  %%%%%%%%%%%%%%%%%%%%%%%%%%%%%%%%%%%%%%%%%%%%%%%%%%%%%
\begin{frontmatter}

\title{Phase-field modelling of solute trapping during rapid
  solidification of a Si--As alloy}

\author{D. Danilov}
\and
\author{B. Nestler}

\address{Institute of Applied Research, Karlsruhe University of
  Applied Sciences, Moltkestrasse 30, 76133 Karlsruhe, Germany}

\date{\empty}

\begin{abstract}
  The effect of nonequilibrium solute trapping by a growing solid
  under rapid solidification conditions is studied using a phase-field
  model.  Considering a continuous steady-state concentration profile
  across the diffuse solid-liquid interface, a new definition of the
  nonequilibrium partition coefficient in the phase-field context is
  introduced. This definition leads, in particular for high growth
  velocities, to a better description of the available experimental
  data in comparison with other diffuse interface and sharp-interface
  predictions.
\end{abstract}

\begin{keyword}
  Rapid solidification; Interface segregation; Bulk diffusion;
  Interface diffusion; Phase-field models.
\end{keyword}

\end{frontmatter}
%%%%%%%%%%%%%%%%%%%%%%%%%%%%%%%%%%%%%%%%%%%%%%%%%%%%%%%%%%%%%%%%%%%%%%

%%%%%%%%%%%%%%%%%%%%%%%%%%%%%%%%%%%%%%%%%%%%%%%%%%%%%%%%%%%%%%%%%%%%%%
%%%  INTRODUCTION  %%%%%%%%%%%%%%%%%%%%%%%%%%%%%%%%%%%%%%%%%%%%%%%%%%%
\section{Introduction}

Experimental results for binary alloys show that under nonequilibrium
conditions, at high growth velocities, impurity concentration could
exceed conventional solid solubility limits given by the equilibrium
phase diagram.  This effect in rapid solidification has been termed
solute trapping and it has been studied extensively with experimental,
theoretical and numerical methods.

\begin{figure}[t]
  \centering
  \includegraphics[angle=-90,width=\textwidth]{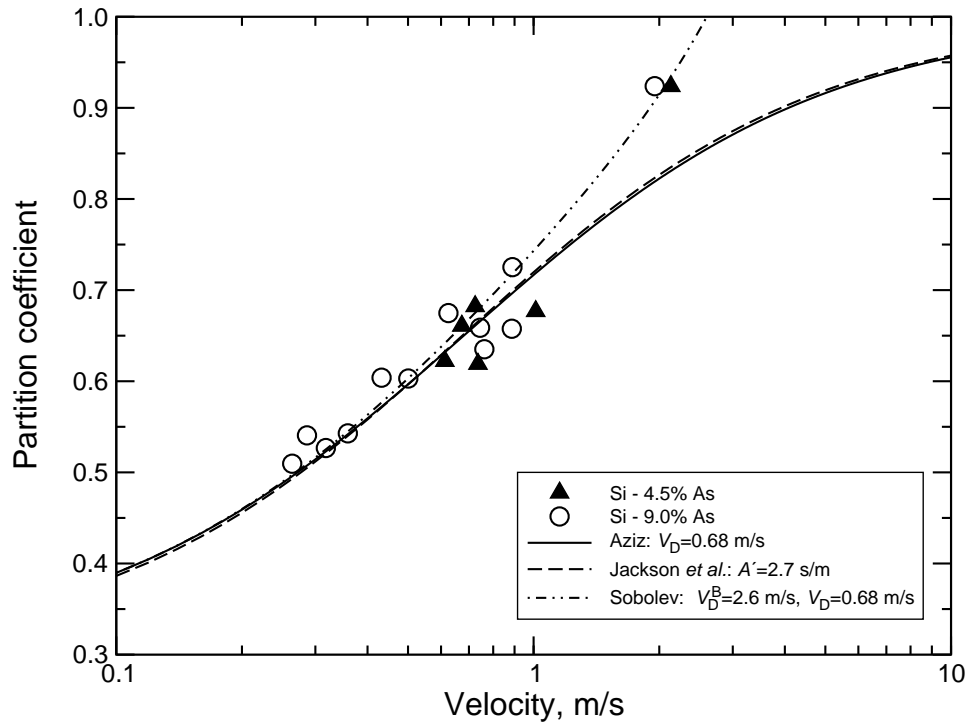}
  \caption{Experimental data \cite{ref:2} ($\blacktriangle$, $\circ$)
    of the nonequilibrium partition coefficient vs. the interface
    velocity for two different binary Si--As alloys in comparison with
    diverse analytical sharp interface models of solute trapping
    (Eqs.~(\ref{eq:2})--(\ref{eq:4})) assuming an equilibrium
    partition coefficient of $k_e=0.3$ \cite{ref:12}.}
  \label{fig:1}
\end{figure}

In the sharp interface approach, a discontinuity in the impurity
concentration from $c_S$ (in the solid) to $c_L$ (in the liquid) at
the interface is assumed, and a partition coefficient is defined by
the ratio
\begin{equation}
  \label{eq:1}
  \text{(sharp-interface) partition coefficient} =
  \frac{\text{concentration in solid}}{\text{concentration in
      liquid}}
  = \frac{c_S}{c_L}.
\end{equation}
According to experimental results for doped silicon
\cite{ref:1,ref:2,ref:3}, the partition coefficient increases from its
equilibrium value $k_e$ and approaches unity as the interface velocity
increases. In these pulsed laser melting experiments, the values of
the partition coefficient are not directly measurable, but they can be
determined by theoretically fitting experimental dopant profiles. The
experimental results \cite{ref:2} for Si--As alloys are reproduced in
Fig.~\ref{fig:1}.  The same transition from an impurity segregation at
low growth velocities to a solute trapping effect in rapid
solidification has been numerically modelled by molecular dynamics
\cite{ref:4} and Monte Carlo \cite{ref:5} methods.

Diverse analytical sharp interface models
\cite{ref:6,ref:7,ref:8,ref:9,ref:10} have been suggested to describe
this phenomenon. Aziz proposed a continuous growth model
\cite{ref:7,ref:8} giving a velocity dependent partition coefficient
of the form
\begin{equation}
  \label{eq:2}
  \frac{c_S}{c_L} = k_A(V) = \frac{k_e + V/V_D}{1 + V/V_D},
\end{equation}
where $V$ is the interface velocity and $V_D$ is a diffusion speed
corresponding to the interface. This form of nonequilibrium partition
coefficient allows to fit experimental data at low and moderate
interface velocities $V < 1$~m/s with $V_D=0.68$~m/s. In the
high-velocity regime at $V\simeq 2$~m/s, the experimental data for the
two Si--As alloys in Fig.~\ref{fig:1} show a much steeper profile than
Eq.~(\ref{eq:2}) proposes. This behaviour is also supported by the
results of a comparison of the predictions of Eq.~(\ref{eq:2}) with
the molecular dynamics simulations (see Fig.~7 of Ref.~\cite{ref:4}).

Jackson \textit{et al.} derived an analytical model \cite{ref:10}
for nonequilibrium segregation leading to the partition coefficient
\begin{equation}
  \label{eq:3}
  \frac{c_S}{c_L} = k_J(V) = k_e^{1/(1+A'V)},
\end{equation}
where the constant $A'$ depends on the diffusion coefficient and on
the interatomic spacing. For $A'=2.7$~s/m, this model also fits well
experimental data and Monte Carlo simulations up to moderate
velocities and exhibits a less steeper profile for rapid
solidification conditions. This can be seen in Fig.~\ref{fig:1} of the
present work and in Fig.~2 of Ref.~\cite{ref:5}. It should be noted
that the best fits of Eqs.~(\ref{eq:2}) and (\ref{eq:3}) to the
experimental data show almost indistinguishable profiles and therefore
the comparison in Section~\ref{sec:3} is given only to
Eq.~(\ref{eq:2}).

Sobolev proposed an extended version \cite{ref:9} of the formula of
Aziz taking into account local nonequilibrium effects in the diffusion
field discussed in Ref.~\cite{ref:11}. The nonequilibrium partition
coefficient reads
\begin{equation}
  \label{eq:4}
  \frac{c_S}{c_L} = k_S(V) =
  \begin{cases}
  \displaystyle\frac{k_e(1-(V/V_D^B)^2) + V/V_D}{1-(V/V_D^B)^2 +
    V/V_D}, & V < V_D^B, \\
  1, & V \geqslant V_D^B,
  \end{cases}
\end{equation}
where $V_D^B$ is a diffusion speed in the bulk liquid. This additional
parameter allows one to fit experimental data in the whole range of
the interface velocities (Fig.~\ref{fig:1}), and predicts the
transition to a completely partitionless solidification with $c_S=c_L$
in Si--As at a finite velocity $V_D^B=2.6$~m/s.

\begin{figure}[t]
  \centering
  \includegraphics[angle=-90,width=\textwidth]{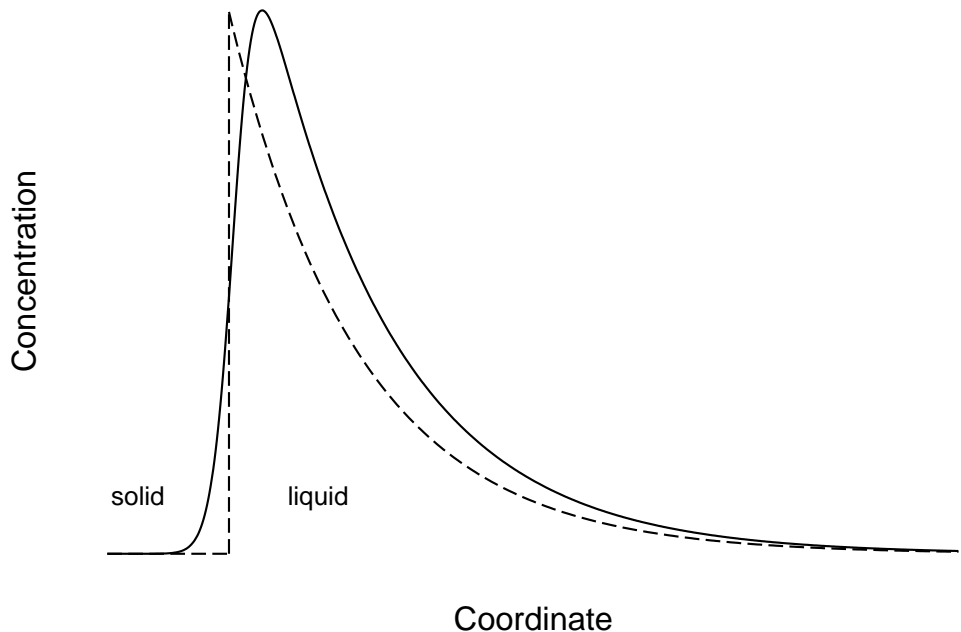}
  \caption{Discontinuous sharp interface concentration profile (dashed
    line) and continuous diffuse interface concentration profile
    (solid line) across a moving solid--liquid interface.}
  \label{fig:11}
\end{figure}

The sharp interface description involves solving diffusion equations
in the phases and joining their solutions at a hypothetical moving
interface of infinitesimally small thickness. However, a transition
interfacial region of finite thickness physically always exists
between the solid and liquid phases. This fact and the effect of the
interaction between the interfacial zone and the diffusion profile in
the liquid (with small diffusion length scale comparable to the
interface thickness at high growth velocities) are included in the
sharp-interface formulation in an implicit way by a separate
consideration of the interface and of the bulk processes.  During the
last two decades, phase-field approaches such as those of
Refs.~\cite{ref:13,ref:14} have extensively been developed to describe
phase transitions. A phase-field model considers in a more realistic
way the diffuse character of the interface with a finite thickness and
describes the dynamical phenomena in both the bulk phases and the
interface region in terms of a single formalism. Much finer scales of
the system are resolved taking into account details of the
concentration profile in the interfacial region. The concentration
profile becomes a continuous function of the coordinate. The specific
jump in the concentration at the interface which is typical for a
sharp interface formulation disappears and is replaced by a continuous
profile with a characteristic maximum near the transition region.
Fig.~\ref{fig:11} exemplarily illustrates the concentration profiles
of sharp interface (dashed line) and diffuse interface (solid line)
models. During the solidification of alloys, solute is rejected ahead
of the growing solid due to the smaller solubility in the solid phase
for partition coefficients $k_e<1$ (see e.g. Ref.~\cite{ref:131}). In
steady-state growth, a concentration boundary layer is established as
a result of the diffusion of atoms.

In the case of a diffuse interface, there is no possibility to relate
a boundary between the solid and liquid phase with a determined
coordinate point and to define further the concentration in the phases
adjacent to the interface (in analogy to sharp-interface models).
Therefore, the definition of the partition coefficient in
Eq.~(\ref{eq:1}) is inapplicable in a phase-field approach and needs
revision.  Wheeler \textit{et al.} \cite{ref:15} and later Ahmad
\textit{et al.} \cite{ref:16} considered rapid solidification and
solute trapping in terms of a phase-field model and suggested the
following definition:
\begin{equation}
  \label{eq:5}
  \text{(phase-field) partition coefficient} = \frac{\text{far-field
    concentration}}{\text{maximum of the concentration}}.
\end{equation}
The definition in Eq.~(\ref{eq:5}) is valid only for steady-state
growth conditions, when the concentration in the bulk solid is equal
to the far-field concentration in the liquid. Applied to the
parameters of a Ni--Cu alloy, the partition coefficients in
Eqs.~(\ref{eq:2}) and (\ref{eq:5}) exhibit a good agreement in the
whole range of interface velocities (see Fig.~2 of
Ref.~\cite{ref:16}).  Using this approach, solute trapping has been
further studied in a diffuse interface model by Kim \textit{et al.}
\cite{ref:170,ref:17} and with numerical methods by Conti
\cite{ref:171}. A rigorous mathematical investigation of the
interfacial conditions and of the solute trapping is provided by
Glasner \cite{ref:172}.

The purpose of this paper is to examine the concentration profiles and
corresponding solute trapping effects at a moving planar interface
using a thermodynamically consistent phase-field model. We compare the
results with available experimental data \cite{ref:2} for a
nonequilibrium partition coefficient in Si--As alloy.  We focus on the
high growth velocity regime, $V>1$~m/s, because for low and
intermediate velocities different models already describe the
experimental data with equal success. A definition of the partition
coefficient at the diffuse interface is suggested predicting a steeper
profile for high growth velocities, $V>1$~m/s, in accordance with the
experimentally measured data of the nonequilibrium partition
coefficient.

%%%%%%%%%%%%%%%%%%%%%%%%%%%%%%%%%%%%%%%%%%%%%%%%%%%%%%%%%%%%%%%%%%%%%%
%%%  MODEL  %%%%%%%%%%%%%%%%%%%%%%%%%%%%%%%%%%%%%%%%%%%%%%%%%%%%%%%%%%
\section{Phase-field model}

We use the phase-field formulation for alloy solidification that has
recently been proposed in Refs.~\cite{ref:18,ref:19} for a general
class of multicomponent and multiphase systems.  The model is based on
an entropy functional and the evolution equations are derived to be
consistent with the first (positive local entropy production) and
second (conservation equations) laws of thermodynamics. The general
case of multiple components is reduced to two phases (solid and
liquid) and to a binary alloy with components A (solvent) and B
(solute).  We apply typical assumptions often made in theories of
solidification processes in binary systems: an ideal solution and
isothermal approximation. A constant temperature $T$ is considered as
a parameter and the free energy density $f$ is postulated in the form
\begin{multline}
  \label{eq:6}
  f(c_A, c_B, \varphi) = \frac{RT}{v_m}
  \left(
    c_A \ln c_A + c_B \ln c_B
  \right) \\
  + \frac{RT}{v_m} \left[
    c_A \ln\left( \frac{1+(T_A-T)/m_e}{1+k_e(T_A-T)/m_e} \right)
    - c_B \ln k_e
  \right] h(\varphi),
\end{multline}
where $T_A$ is the melting point of the pure component A, $c_A$ and
$c_B$ are the concentrations of the alloy components given in molar
fractions, $k_e$ and $m_e$ are the partition coefficient and the
liquidus slope of the equilibrium phase diagram, $v_m$ is the molar
volume and $R$ is the gas constant. The phase-field variable $\varphi$
describes the thermodynamic state of a local volume. The value
$\varphi=1$ corresponds to the solid phase, $\varphi=0$ corresponds to
the liquid and the function $h(\varphi)=\varphi^2(3-2\varphi)$ is
monotonic in the interval $[0,1]$ satisfying the conditions $h(0)=0$
and $h(1)=1$ in the bulk phases. The form of the free energy density
in Eq.~(\ref{eq:6}) leads to an equilibrium phase diagram with
straight solidus $T_S=T_A + m_e c_B / k_e$ and liquidus $T_L=T_A + m_e
c_B$ lines by a conventional common tangent construction.

Considering a planar solid-liquid interface, we write the evolution
equations of the phase-field model for isotropic kinetics and
isotropic surface energies of the interface. The evolution of the
phase-field variable $\varphi$ is determined by the partial
differential equation
\begin{equation}
  \label{eq:7}
  \frac{2 \varepsilon\gamma}{\nu} \frac{\partial \varphi}{\partial t}
  = 2 \varepsilon \gamma \frac{\partial^2 \varphi}{\partial x^2}
  - \frac{9 \gamma}{\varepsilon} \frac{\partial g(\varphi)}{\partial \varphi}
  - \frac{1}{T} \frac{\partial f}{\partial \varphi},
\end{equation}
where $g(\varphi)=\varphi^2(1-\varphi)^2$ is a double well potential,
$\gamma$ is the entropy density of the solid--liquid interface, $\nu$
is the interface mobility and the parameter $\varepsilon$ determines
the thickness of the interfacial zone. The concurrence between the
first and the second terms on the right-hand side of Eq.~(\ref{eq:7})
generates the diffuse transition zone between the phases. The last
term drives the growth.

The diffusion mass transport of the alloy component B is determined by
the nonlinear diffusion equation
\begin{equation}
  \label{eq:8}
  \frac{\partial c_B}{\partial t}
  = \frac{\partial}{\partial x} \left(
    D(\varphi) \frac{\partial c_B}{\partial x}
  \right)
  - \Theta \frac{\partial}{\partial x}
  \left(
    D(\varphi) c_B(1-c_B) \frac{\partial h}{\partial x}
  \right),
\end{equation}
where $D(\varphi)=\varphi D_S + (1-\varphi) D_L$ with constant
diffusion coefficients $D_S$ and $D_L$ of component B in the solid and
in the liquid phase, respectively. To derive Eq.~(\ref{eq:8}), the
constraint condition $c_A+c_B=1$ has been applied. The quantity
$\Theta$ denotes the driving force for the redistribution of the alloy
components at the solid--liquid interface:
\begin{equation}
  \label{eq:9}
  \Theta
  = \ln \left(
    \frac{1+(T_A-T)/m_e}{1+k_e(T_A-T)/m_e}
  \right)
  + \ln(k_e).
\end{equation}

Growth far from the equilibrium is usually accompanied by a
steady-state motion of the interface leading to steady-state
concentration profiles. Thus it is useful to consider the velocity $V$
and the solute concentration $c_0$ in the liquid far from the
interface as control parameters whereas the self-consistent
temperature and the interfacial concentration will be determined by
Eqs.~(\ref{eq:7}) and (\ref{eq:8}).  Looking for a steady-state
solution, we adopt a frame of reference
\begin{equation}
  \label{eq:10}
  z = x - Vt,
\end{equation}
propagating at a constant velocity $V$ and coincident with the centre
of the interfacial zone given by $\varphi=1/2$ at $z=0$.
Eqs.~(\ref{eq:7}) and (\ref{eq:8}) read
\begin{equation}
  \label{eq:11}
  - V \frac{2 \varepsilon\gamma}{\nu} \frac{\partial \varphi}{\partial z}
  = 2 \varepsilon \gamma \frac{\partial^2 \varphi}{\partial z^2}
  - \frac{9 \gamma}{\varepsilon} \frac{\partial g}{\partial \varphi}
  - \frac{1}{T} \frac{\partial f}{\partial \varphi},
\end{equation}
\begin{equation}
  \label{eq:12}
  -V \frac{\partial c_B}{\partial z}
  = \frac{\partial}{\partial z} \left(
    D(\varphi) \frac{\partial c_B}{\partial z}
  \right)
  - \Theta \frac{\partial}{\partial z}
  \left(
    D(\varphi) c_B(1-c_B) \frac{\partial h}{\partial z}
  \right).
\end{equation}
To investigate the dependence of the concentration profile and the
system temperature on the growth velocity, we have computed the
numerical solutions of the nonlinear Eqs.~(\ref{eq:11}) and
(\ref{eq:12}) with associated boundary conditions
\begin{equation}
  \label{eq:13}
   \begin{gathered}
    \varphi|_{z \rightarrow\, -\infty} = 1,
    \quad
    \varphi|_{z \rightarrow\, +\infty} = 0, \\
    \frac{\partial c_B}{\partial z}|_{z \rightarrow\, -\infty} = 0,
    \quad
    c_B|_{z \rightarrow\, +\infty} = c_0.
  \end{gathered}
\end{equation}
The spatial derivatives have been discretised using finite differences
on a uniform grid. A relaxation method is used to obtain the solution
of the nonlinear system in Eqs.~(\ref{eq:11})--(\ref{eq:13}). Starting
from an initial guess ($\varphi^0(z)$, $c^0(z)$, $T^0$), the
successive iterations ($\varphi^n(z)$, $c^n(z)$, $T^n$) for
$n\geqslant1$ are computed until the convergence criteria
$||\varphi^{n+1}-\varphi^{n}||<10^{-9}$ and
$||c^{n+1}-c^{n}||<10^{-9}c_0$ are reached, where $||\psi||$ is the
$L_2$-norm of the finite difference representation of a function
$\psi(z)$. To ensure smoothness of the numerical solution in the
interfacial region, the grid resolution $\Delta x$ has been chosen
from $\Delta x = \frac{1}{20}\varepsilon$ at low velocities to $\Delta
x = \frac{1}{80}\varepsilon$ at high velocities.

%%%%%%%%%%%%%%%%%%%%%%%%%%%%%%%%%%%%%%%%%%%%%%%%%%%%%%%%%%%%%%%%%%%%%%
%%%  RESULTS  %%%%%%%%%%%%%%%%%%%%%%%%%%%%%%%%%%%%%%%%%%%%%%%%%%%%%%%%
\section{Results}
\label{sec:3}

In this section, we consider the application of the model in
Eqs.~(\ref{eq:11}) and (\ref{eq:12}) to solidification of a Si--As
alloy. The thermophysical parameters of the alloy are contained in the
diffusion equation (diffusion coefficients) and in the entropy
contributions: the double well potential, the gradient and the free
energy density functions. The values used in the computations are
listed in Table~\ref{tab:params}.  The liquidus slope $m_e$ has been
estimated from the phase diagram given in Ref.~\cite{ref:20}.  The
value of the interface mobility $\nu=1.22\times 10^{-8}$~m$^2$/s has
been adjusted to match the kinetic coefficient $\Delta
T/V=15$~(K$\cdot$s)/m at high undercoolings $\Delta T$ as assumed in
Ref.~\cite{ref:2}. The work in Refs.~\cite{ref:15,ref:16} shows that
the magnitude of the solute trapping effect is directly related to the
phase-field parameter $\varepsilon$, and a comparison of
large-velocity expansions of Eqs.~(\ref{eq:2}) and (\ref{eq:5}) leads
to a relation $V_D \sim D/\varepsilon$ between diffusion speed $V_D$,
diffusion coefficient $D$, and parameter $\varepsilon$. On this basis,
we choose the value of $\varepsilon=3.5\times 10^{-9}$~m by fitting
the phase-field partition coefficient in Eq.~(\ref{eq:5}) to the sharp
interface partition coefficient in Eq.~(\ref{eq:2}) using the value
$V_D=0.68$~m/s provided by comparison with the experimental data
(Fig.~\ref{fig:1}). By this selection of the parameter $\varepsilon$
on the physical scale of nanometers no correction terms have been
introduced into the model. The inclusion of correction terms such as
thin-interface asymptotics in Refs.~\cite{ref:201,ref:203} or
anti-trapping term in Refs.~\cite{ref:202,ref:203} is needed for
applications to model dendritic growth and microstructure evolution
processes in two and three dimensions at low undercoolings when the
interface thickness is some orders of magnitude larger than
nanometers. The far-field concentration in the melt has been set to
$c_0=0.09$ reflecting the alloy composition Si--9\,at.\%\,As.

\begin{table}[b]
  %\centering
  \caption{Physical parameters of Si--9\,at.\%\,As alloy used for the
    phase-field simulations.}
  \label{tab:params}
  \begin{tabular}{ccc}
    \hline
    Parameter & Value & Reference \\
    \hline
    $T_A$ & $1685$~K & \cite{ref:21} \\
    $m_e$ & $-400$~K &  \\
    $k_e$ & $0.3$    & \cite{ref:12} \\
    $v_m$ & $12\times 10^{-6}$~m$^3$  &  \\
    $D_L$ & $1.5\times 10^{-9}$~m$^2$/s    & \cite{ref:2} \\
    $D_S$ & $3\times 10^{-13}$~m$^2$/s    & \cite{ref:2} \\
    $\gamma$ & $2.8\times 10^{-4}$~J/(K$\cdot$m$^2$)    & \cite{ref:21} \\
    $\nu$ & $1.22\times 10^{-8}$~m$^2$/s  &  \\
    $\varepsilon$ & $3.5\times 10^{-9}$~m  &  \\
    \hline
  \end{tabular}
\end{table}

\begin{figure}[t]
  \centering
  \includegraphics[angle=-90,width=\textwidth]{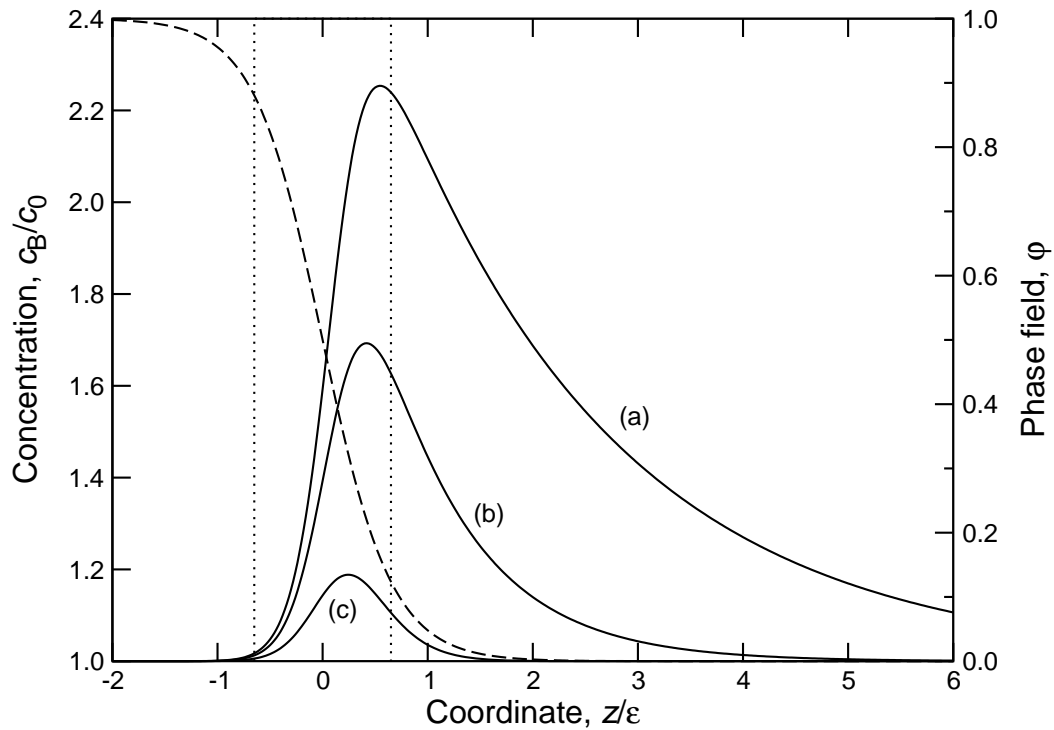}
  \caption{Steady-state concentration profiles (solid lines) for three
    different velocities: (a) $V=0.2$~m/s, (b) $V=0.5$~m/s, (c)
    $V=2$~m/s, and phase-field profile (dashed line) in a reference
    frame moving in a steady-state manner with the interface.}
  \label{fig:2}
\end{figure}

Fig.~\ref{fig:2} shows the concentration profiles (solid lines) for
three different values of the interface velocity. The concentration
profile in the solid phase has a uniform value equal to the far-field
concentration $c_0$ in the liquid. In the interfacial region, the
concentration increases due to a rejection of solute atoms by the
growing solid. In the liquid ahead of the interface, a concentration
boundary layer is formed by diffusion transport of the rejected solute
atoms into the liquid.  With increasing velocity the inhomogeneity of
the concentration field reduces. Both the maximum value of the solute
concentration as well as the spatial penetration of the concentration
profile into the liquid diminish indicating a reduction of solute
segregation at the interface as characteristic for the occurrence of
solute trapping. The phase-field profile (dashed line) exhibits only a
weak dependence on the interface velocity.

\begin{figure}[t]
  \centering
  \includegraphics[angle=-90,width=\textwidth]{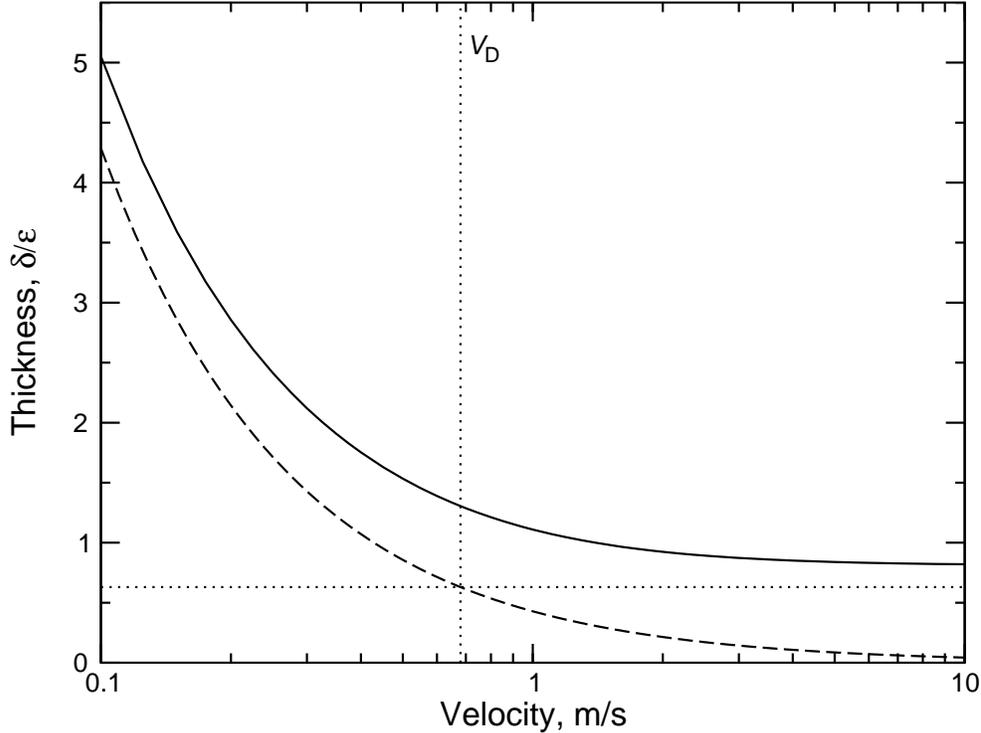}
  \caption{The dependence of the characteristic length scale $\delta$
    (solid line) and $\delta_{SI}$ (dashed line) on the interface
    velocity.}
  \label{fig:3}
\end{figure}

To analyse quantitatively the concentration profile, we introduce a
characteristic length scale $\delta$ describing its thickness by the
definition
\begin{equation}
  \label{eq:14}
  \int_{-\infty}^{+\infty} \big(c_B(z)-c_0\big) dz
  = \big(c_m-c_0\big) \delta,
\end{equation}
where $c_m$ is the maximum of the concentration.  From
Eq.~(\ref{eq:14}), the corresponding sharp interface steady-state
concentration profile \cite{ref:22}
\begin{equation}
  \label{eq:15}
  \bar{c}(z) =
  \begin{cases}
    c_0 + (c_L-c_0)
    \exp\bigg(
      -\displaystyle \frac{Vz}{D_L}
    \bigg), & z > 0, \\
    c_0, & z < 0,
  \end{cases}
\end{equation}
leads to the characteristic length scale
\begin{equation}
  \label{eq:16}
  \delta_{SI} = \frac{D_L}{V},
\end{equation}
where the maximum of the concentration $c_m$ corresponds to the
interface concentration $c_L$ at the side of the liquid phase.  The
dependence of the length scales $\delta$ and $\delta_{SI}$ on the
interface velocity $V$ is shown in Fig.~\ref{fig:3}.  At low interface
velocities, both $\delta$ and $\delta_{SI}$ are larger than the
interface length scale $\varepsilon$. At high velocities, the
thickness $\delta$ of the phase-field solution (solid line) approaches
a constant value comparable to $\varepsilon$, whereas the thickness
$\delta_{SI}$ of the sharp interface profile (dashed line) tends to
zero.  The dotted lines mark the position where the velocity $V$
equals the diffusion speed $V_D$, i.e. $V=V_D$ and
$\delta_{SI}=0.6\varepsilon$. This position can be considered as a
limit for the range of validity of the sharp interface description.
At velocities $V>V_D$, the finite thickness of the solid-liquid
interface cannot be ignored in the description of the diffusion
transport of rejected B-atoms away from the moving interface.  The
tendency of the length scale $\delta$ to become constant as the
interface velocity increases is accompanied by a shift of the position
of the concentration maximum to the centre of the interfacial zone.
Hence, under rapid solidification conditions, the inhomogeneity of the
concentration field is completely contained in the diffuse interface
region and the concentration field in the bulk liquid tends to the
uniform value of the far-field concentration $c_0$
(Fig.~\ref{fig:2}\,(c)).

Considering Fig.~\ref{fig:2}, two definitions of the partition
coefficient at the diffuse interface are obvious.  First, following
Wheeler \textit{et al.} \cite{ref:15}, the solute concentration in
the liquid at the interface is associated with the maximum of the
concentration field $c_m$ and the concentration in the solid is
assumed to be equal to $c_0$. This leads to a partition coefficient
$k_m$ that reads
\begin{equation}
  \label{eq:17}
  k_m = \frac{c_0}{c_m}.
\end{equation}

Second, in the present work we suggest a definition where the
appropriate solid and liquid concentrations for the partition
coefficient are determined at positions $z_S$ and $z_L$ in the
interior of the diffuse interfacial region leading to the expression
\begin{equation}
  \label{eq:18}
  k = \frac{c(z_S)}{c(z_L)}.
\end{equation}
The positions $z_S$ and $z_L$ can be chosen relative to the centre of
the interfacial zone in a symmetric way with a best fit criterion to
the experimental data.

\begin{figure}[t]
  \centering
  \includegraphics[angle=-90,width=\textwidth]{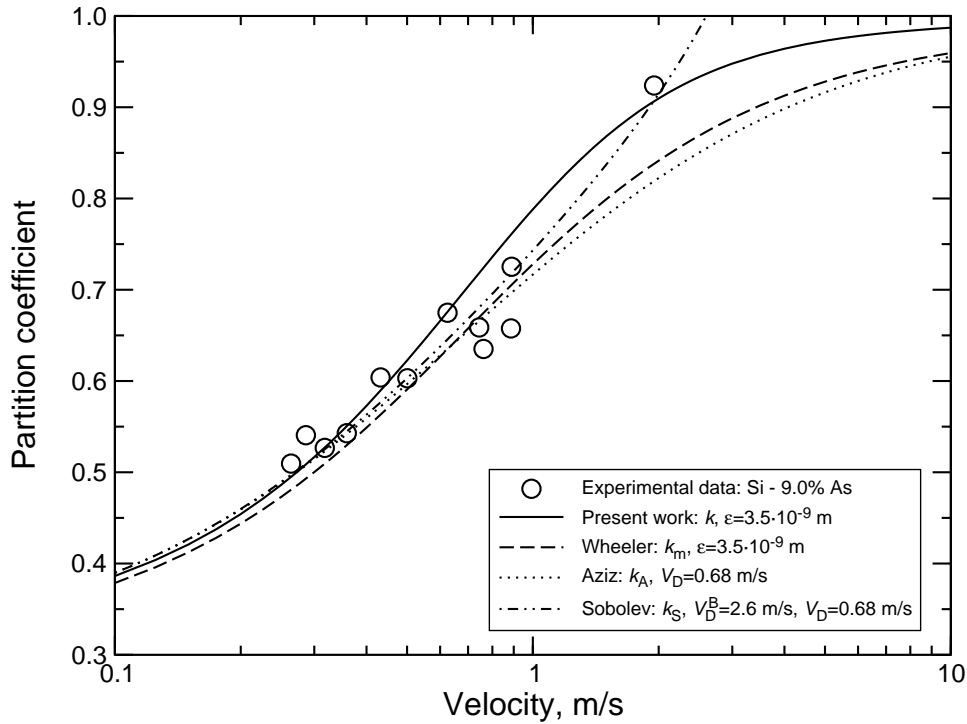}
  \caption{Different models for the partition coefficient as a
    function of the interface velocity applied to the alloy system
    Si--9\,at\%\,As and compared with experimental data taken from
    Ref.~\cite{ref:2}.}
  \label{fig:4}
\end{figure}

Fig.~\ref{fig:4} displays the different behaviour of the partition
coefficients given by Eqs.~(\ref{eq:2}), (\ref{eq:4}), (\ref{eq:17})
and (\ref{eq:18}) as a functions of the interface velocity $V$. The
theoretical models are applied to the alloy system Si--9\,at.\%\,As,
since experimental data are provided in Ref.~\cite{ref:2}.  The
definition $k$ in Eq.~(\ref{eq:18}) has a steeper profile for high
velocities $V > 1$ m/s compared to both the behaviour of $k_m$ in
Eq.~(\ref{eq:17}) of the phase-field model for rapid solidification in
Ref.~\cite{ref:15} and the predictions for $k_A$ in Eq.~(\ref{eq:2})
of the continuous growth model in Refs.~\cite{ref:7,ref:8}.  For the
present data of Si--9\,at.\%\,As, the steeper profile is in better
agreement with the experimentally measured partition coefficient for
$V=2$~m/s.  The curve of $k$ in Fig.~\ref{fig:4} was obtained for the
positions $z_S=-0.65\varepsilon$ and $z_L=0.65\varepsilon$ of the
solid and liquid concentrations, respectively.  The three approaches
for $k(V)$, $k_m(V)$ and $k_A(V)$ show an asymptotic convergence of
the partition coefficient to one for $V \rightarrow \infty$.  The
smaller convergence rate for $k_m(V)$ and $k_A(V)$ can exemplarily be
illustrated by comparing the values where $1-k=1.28\times 10^{-2}$,
$1-k_m=4.06\times 10^{-2}$ and $1-k_A=4.46\times 10^{-2}$ at the
velocity $V=10$~m/s. The value of $1-k_A=1.28\times 10^{-2}$ for the
continuous growth model is reached only at a significantly larger
velocity $V=36.4$~m/s.  In contrast to the asymptotic convergence of
the three approaches for $k(V)$, $k_m(V)$ and $k_A(V)$, the model of
Sobolev predicts a sharp transition to $k_S(V) = 1$ at the velocities
$V\geqslant V_D$. Thus, the experimental data for rapid solidification
can be described by the sharp interface approach in Eq.~(\ref{eq:4})
and by the diffuse interface approach in Eq.~(\ref{eq:18}). However
these models predict a qualitatively different character of the
transition to establish complete solute trapping.

%%%%%%%%%%%%%%%%%%%%%%%%%%%%%%%%%%%%%%%%%%%%%%%%%%%%%%%%%%%%%%%%%%%%%%
%%%  SUMMARY  %%%%%%%%%%%%%%%%%%%%%%%%%%%%%%%%%%%%%%%%%%%%%%%%%%%%%%%%
\section{Summary}

Considering the continuous concentration profiles at the diffuse
solid--liquid interface in Fig.~\ref{fig:2}, the definition in
Eq.~(\ref{eq:18}) of the nonequilibrium partition coefficient in the
phase-field description of solute trapping during rapid solidification
is proposed. The dependence of the partition coefficient on the
interface velocity is compared with other diffuse and sharp interface
model predictions. Under the assumption of local equilibrium in the
bulk phases, for high interface velocities $V>1$~m/s, the expression
in Eq.~(\ref{eq:18}) gives a steeper profile than previous models,
Eqs.~(\ref{eq:2}), (\ref{eq:3}), (\ref{eq:17}), and is here in better
agreement with the available experimental data for Si--9\,at.\%\,As in
the range of high growth velocities.  Due to a lack of experimental
data, a more extended verification of the quantitative character of
the transition to complete solute trapping is open: the question to be
clarified is whether there exist a sharp transition as predicted by
local nonequilibrium solidification model \cite{ref:11} or a smooth
continuous transition as predicted by the continuous growth model
\cite{ref:7,ref:8} and phase-field models \cite{ref:16}.

The presented model of solute trapping can be generalised further by
including the effects of local nonequilibrium in the concentration
field into the phase-field formulation as described by Galenko
\cite{ref:23}. An examination of this effect on the partition
coefficient in the diffuse interface approach will be a task for
future investigations.

%%%%%%%%%%%%%%%%%%%%%%%%%%%%%%%%%%%%%%%%%%%%%%%%%%%%%%%%%%%%%%%%%%%%%%
%%%  ACKNOWLEDGEMENTS  %%%%%%%%%%%%%%%%%%%%%%%%%%%%%%%%%%%%%%%%%%%%%%%
\begin{ack}
  This work was supported by the German Research Foundation (DFG)
  under the priority research program 1120: ``Phase transformations in
  multi-component melts'', Grant No. Ne 822/3-1. The funding is
  gratefully acknowledged.
\end{ack}

%%%%%%%%%%%%%%%%%%%%%%%%%%%%%%%%%%%%%%%%%%%%%%%%%%%%%%%%%%%%%%%%%%%%%%
%%%%  BIBLIOGRAPHY  %%%%%%%%%%%%%%%%%%%%%%%%%%%%%%%%%%%%%%%%%%%%%%%%%%

%%%%%%%%%%%%%%%%%%%%%%%%%%%%%%%%%%%%%%%%%%%%%%%%%%%%%%%%%%%%%%%%%%%%%%
%%%  END  %%%%%%%%%%%%%%%%%%%%%%%%%%%%%%%%%%%%%%%%%%%%%%%%%%%%%%%%%%%%

\begin{thebibliography}{10}

\bibitem{ref:1}
Aziz MJ, Tsao JY, Thompson MO, Peercy PS, White CW. Phys Rev Lett
1986;56:2489.

\bibitem{ref:2}
Kittl JA, Aziz MJ, Brunco DP, Thompson MO. J Cryst Growth
1995;148:172.

\bibitem{ref:3}
Kittl JA, Sanders PG, Aziz MJ, Brunco DP, Thompson MO. Acta Mater
2000;48:4797.

\bibitem{ref:4}
Celestini F, Debierre JM. Phys Rev~B 2000;62:14006.

\bibitem{ref:5}
Beatty KM, Jackson KA. J~Cryst Growth 2004;271:495.

\bibitem{ref:6}
Wood RF. Phys Rev~B 1982;25:2786.

\bibitem{ref:7}
Aziz MJ. J Appl Phys 1982;53:1158.

\bibitem{ref:8}
Aziz MJ, Kaplan T. Acta Metall 1988;36:2335.

\bibitem{ref:9}
Sobolev SL. Phys Stat Sol (a) 1996;156:293.

\bibitem{ref:10}
Jackson KA, Beatty KM, Gudgel KA. J~Cryst Growth 2004;271:481.

\bibitem{ref:11}
Galenko P, Sobolev S. Phys Rev~E 1997;55:343.

\bibitem{ref:12}
Reitano R, Smith PM, Aziz MJ. J Appl Phys 1994;76:1518.

\bibitem{ref:13}
Chen LQ. Annu Rev Mater Res 2002;32:113.

\bibitem{ref:14}
Boettinger WJ, Warren JA, Beckermann C, Karma A. Annu Rev Mater Res
2002;32:163.

\bibitem{ref:131}
Kurz W, Fisher DJ. Fundamentals of solidification. TTP, 1998.

\bibitem{ref:15}
Wheeler AA, Boettinger WJ, McFadden GB. Phys Rev E 1993;47:1893.

\bibitem{ref:16}
Ahmad NA, Wheeler AA, Boettinger WJ, McFadden GB. Phys Rev E 1998;58:3436.

\bibitem{ref:170}
Kim SG, Kim WT, Suzuli T. Phys Rev E 1999;60:7186.

\bibitem{ref:17}
Kim WT, Kim SG. Mater Sci and Eng A 2001;304--306:220.

\bibitem{ref:171}
Conti M. Phys Rev E 1997;56:3717.

\bibitem{ref:172}
Glasner K. Physica D 2001;151:253.

\bibitem{ref:18}
Garcke H, Nestler B, Stinner B. SIAM J Appl Math 2004;64:775.

\bibitem{ref:19}
Nestler B, Garcke H, Stinner B. Phys Rev E 2005;71:041609.

\bibitem{ref:20}
Predel B. Phase Equilibria of Binary Alloys. Springer, 2003.

\bibitem{ref:201}
Karma A, Rappel WJ. Phys Rev E 1996;53:R3017.

\bibitem{ref:202}
Karma A. Phys Rev Lett 2001;87:115701.

\bibitem{ref:203}
Echebarria B, Folch R, Karma A, Plapp M. Phys Rev E 2004;70:061604.

\bibitem{ref:21}
Vinet B, Magnusson L, Fredriksson H, Desr\'e PJ. J Colloid and
Interface Science 2002;255:363.

\bibitem{ref:22}
Ivantsov GP. Dokl Akad Nauk SSSR 1951;81:179.

\bibitem{ref:23}
Galenko P, Phys Lett A 2001;287:190.

\end{thebibliography}
\end{document}